\documentclass[manuscript,nonacm]{acmart}
\AtBeginDocument{%
  \providecommand\BibTeX{{%
    \normalfont B\kern-0.5em{\scshape i\kern-0.25em b}\kern-0.8em\TeX}}}

\setcopyright{rightsretained}
\copyrightyear{2021}
\acmYear{2021}
\acmDOI{}

\usepackage{adjustbox}
\usepackage{tabularx}
\usepackage{float}
\usepackage{amsmath}
\usepackage{breqn}

\newcommand{\cond}{\nonscript\;\middle|\nonscript\;}
\newcommand{\ranking}[0]{L}
\newcommand{\allitems}[0]{D}
\newcommand{\allusers}[0]{Q}

\newcommand{\doc}[0]{d}
\newcommand{\docs}[1]{\doc_{#1}}
\newcommand{\user}[0]{q}
\newcommand{\users}[1]{q_{#1}}

\newcommand{\prefix}[1]{\ranking_{\le {#1}}}

\newcommand{\rankingPosition}[0]{\ranking(\doc)}
\newcommand{\relevance}[0]{y(d|q)}
\newcommand{\predictedrelevance}[0]{\hat{y}(d|q)}

\newcommand{\group}[0]{g}
\newcommand{\groups}[0]{\mathcal{G}}

\newcommand{\progroup}[0]{\groups^+}
\newcommand{\nonprogroup}[0]{\groups^-}

\newcommand{\Protected}[0]{\progroup(\ranking)}
\newcommand{\nonProtected}[0]{\nonprogroup(\ranking)}

\newcommand{\alignmentvec}[0]{\groups(\doc)}
\newcommand{\alignmentmat}[0]{\groups(\ranking)}

\newcommand{\populationEstimator}[0]{\hat{\textbf{p}}}
\newcommand{\probabiltyDist}[0]{\hat p}

\newcommand{\attention}[0]{\mathbf{a}}
\newcommand{\attentionvec}[0]{\attention_{\ranking}(\doc)}
\newcommand{\attentionmat}[0]{\attention_{\ranking}}

\newcommand{\exposure}[0]{\boldsymbol\epsilon}
\newcommand{\groupexposure}[0]{\exposure_{\ranking}}

\newcommand{\stochasticRanking}[0]{\pi}

\newcommand{\prefd}[0]{\mathrm{PreF}_\Delta}
\newcommand{\AWRF}[0]{\mathrm{AWRF}_\Delta}

\newcommand{\expectation}[2]{\operatorname{E}_{#1}[#2]}
\begin{document}

\title{Comparing Fair Ranking Metrics}

\author{Amifa Raj}
\email{amifraj@u.boisestate.edu}
\affiliation{%
  \institution{People and Information Research Team, Boise State University}
  \city{Boise}
  \state{Idaho}
  \postcode{83725-2055}
}
\author{Michael D. Ekstrand}
\email{michaelekstrand@boisestate.edu}
\affiliation{%
  \institution{People and Information Research Team, Boise State University}
  \city{Boise}
  \state{Idaho}
  \postcode{83725-2055}
}

\renewcommand{\shortauthors}{Raj et al.}

\begin{abstract}
Ranked lists are frequently used by information retrieval (IR) systems  to present results believed to be relevant to the user's information need. 
\textit{Fairness} is a relatively new but important aspect of these rankings to measure, joining a rich set of metrics that go beyond traditional accuracy or utility constructs to provide a more holistic understanding of IR system behavior.
In the last few years, several metrics have been proposed to quantify the (un)fairness of rankings, particularly with respect to particular group(s) of content providers, but comparative analyses of these metrics --- particularly for IR --- is lacking.
There is limited guidance, therefore, to decide what fairness metrics are applicable to a specific scenario, or assessment of the extent to which metrics agree or disagree applied to real data.
%We aim to bridge the gap between theoretical and practical application of these metrics.
In this paper, we describe several fair ranking metrics from existing literature in a common notation, enabling direct comparison of their assumptions, goals, and design choices; we then empirically compare them on multiple data sets covering both search and recommendation tasks.
%We also provide a sensitivity analysis to assess the impact of the design choices and parameter settings that go in to these metrics and point to additional work needed to improve fairness measurement.

\end{abstract}

%Ranking is a fundamental aspect of recommender systems. However, ranked outputs can be susceptible to various biases; some of these may cause disadvantages to members of protected groups. Several metrics have been proposed to quantify the (un)fairness of rankings, but there has not been to date any direct comparison of these metrics. This complicates deciding what fairness metrics are applicable for specific scenarios, and assessing the extent to which metrics agree or disagree. In this paper, we describe several fair ranking metrics in a common notation, enabling direct comparison of their approaches and assumptions, and empirically compare them on the same experimental setup and data set. Our work provides a direct comparative analysis identifying similarities and differences of fair ranking metrics selected for our work.
%\end{abstract}



\maketitle
\section{Introduction}

Ranked lists are frequently used in information retrieval (IR) to present items in response to users' information needs.
% Ranking algorithms order items based on relevance scores (and sometimes other measures of items' quality and relationships, such as similarity in maximal marginal relevance \cite{carbonell1998use}).
Through the ranked list, a system \textit{exposes} items (and their providers) to users, and this visibility affects what users discover, consume, and purchase. Further, exposure is not always evenly or fairly distributed \cite{diaz2020evaluating}. Inequitable exposure can disadvantage content providers individually (when an item receives more or less exposure than other items of comparable relevance) or on a group basis (when a group of items or providers are systematically under- or over-exposed). Fairness in IR is a broad topic encompassing many concerns, including the concerns of multiple stakeholders \citep{burke:multisided} and variety of fairness objectives that may be pursued \citep{zehlike2021fairness}.
Fairness is also a complex and contested social construct \citep{selbst:abstraction}, so there is not one fairness goal that can be achieved; rather, fairness objectives and metrics need to be selected for a particular application and set of concerns.
For a fuller treatment of fairness, \citet{mitchell2021algorithmic} provide an overview of general concepts of fairness in machine learning systems, and \citet{ekstrand2021fairness} systematize fairness in IR.
%The system may also, however, provide greater or lesser exposure reflecting in a ways that reproduce historical and ongoing social discrimination, such as by gender or race. 
%
%Popularity bias \cite{abdollahpouri2019popularity}, for example, provides an advantage to creators based on their prior popularity. Providers whose work is already widely-consumed are likely to be more frequently recommended in the future, particularly by collaborative filters \cite{abdollahpouri2019unfairness}.

In the last few years, several metrics have been proposed to measure the \textit{fairness} of rankings in this sense, along with various goals for what it means for a ranking to be fair or unfair towards items and their providers. 
% some directly in IR contexts, and others for more general ranking purposes such as college rankings or university admissions.
These metrics have typically been tested on a different applications and data sets, some in IR but many in the contexts of  college rankings or university admissions. \citet{kuhlman2021measuring} compare selected fair ranking metrics for measuring the \emph{statistical parity} of rankings (whether they provide equal exposure to different groups), and \citet{zehlike2021fairness} provides a thorough conceptual survey of fair ranking constructs, but there has not yet been a systematic comparison of group fairness metrics for ranked IR outputs (where the system provides different rankings in response to for different information needs --- both prior comparisons focus on rankings for a single need), or direct comparisons within the same data set and experiment.
Applying fair ranking metrics to real IR experiment data reveals challenges with their practical applications, including incomplete data (for both relevance and group membership) and the occurrence of edge cases such as groups with no relevant (or retrieved) items.
% Metrics need to be robust and usable in such situations in order to be practically useful in experiments and for auditing deployed applications.

In this work we specifically consider ways of assessing if the rankings a system produces are fair or unfair to content providers from particular (often demographic) groups; this is \textit{provider-side group fairness} \citep{burke:multisided}. We adopt the common frame inspired by United States anti-discrimination law of a ``protected group'': a class of people who share a trait upon which a retrieval or classification should not be discriminatory \citep{xiangandraji}. Our goal is to provide insight on how to measure the provider-side group fairness of the ranked outputs in actual IR experiments (search and recommendations) using metrics from the existing literature, comparing them and documenting limitations in their practical application. 
This paper has three contributions:
\begin{itemize}
    \item We describe rank fairness metrics in a unified notation for IR applications, identifying design points, similarities, and differences.
    \item We identify gaps between the conceptual form of the metrics and the practicalities of applying them search and recommendation experiments.%\footnote{Code will be published upon acceptance.}
    \item We directly compare metric results with the same data and systems across multiple search and recommendation data sets.
    %\item We conduct sensitivity analysis to assess the impact of design choices and external factors on these metrics.
\end{itemize}

\section{Fair Ranking Metrics}
\label{sec:metrics}
\begin{table*}[tbh]
%\centering
\caption{Summary of fair ranking metrics.}
\label{tbl:summary}
% \begin{adjustbox}{\textwidth}
\begin{minipage}{\textwidth}
\centering

\begin{tabular}{lccccc}
Metric(s) & Weighting & Target & Binomial? & Range & More Fair\\
\toprule
$\prefd$ \citep{yang2017measuring} & --- & $\hat p$ from full ranking & Dep. on $\Delta$\footnote{$\Delta_{\mathrm{RD}}$ and $\Delta_{\mathrm{RD}}$ both use binomial protected group attributes, but $\Delta_{\mathrm{KL}}$ generalizes.} & $[0,1]$ & 0 \\
$\AWRF$ \citep{sapiezynski2019quantifying} & Geometric & configured $\hat p$ & Dep. on $\Delta$ & $[0,1]$ & 0 \\
FAIR \citep{zehlike2017fa}  & --- & binomial $\hat p$ & Yes & $[0,1]$ & 1 \\
\midrule
logDP \citep{singh2018fairness} & Logarithmic & equality & Yes & $(-\infty,\infty)$ & 0\\
logEUR \citep{singh2018fairness} & Logarithmic & $\propto \text{utility}$ & Yes & $(-\infty,\infty)$ & 0 \\
logRUR \citep{singh2018fairness} & Logarithmic & $\propto \text{disc. utility}$ & Yes & $(-\infty,\infty)$ & 0\\
IAA \citep{biega2018equity} & Geometric & $\propto \text{est. utility}$ & No & $[0,\infty)$ & 0\\
EEL, EER \citep{diaz2020evaluating} & Cascade, RBP. & $f(\text{utility})$ & No & $[0,\infty)$ & EEL 0, EER $>$\\
EED \citep{diaz2020evaluating} & Cascade\footnote{Cascade weighting also incorporates relevance into exposure, even if exposure is not compared to relevance.}, RBP. & equality & No & $[0,\infty)$ & 0 \\
\midrule
PAIR \citep{beutel2019fairness, narasimhan2020pairwise} & --- & equal accuracy & No & $[0,1]$ & 0 \\
\bottomrule
\end{tabular}
\end{minipage}
% \end{adjustbox}
\end{table*}
We begin by describing the fair ranking metrics, summarized in table~\ref{tbl:summary}, in a common framework and notation.
This enables direct comparison of their designs and theoretical behavior, and facilitates easier implementation in IR experiments. In some cases, we assign new name for metrics based on their functionality, purpose, and comparability within our synthesis.
%Some constructs assess fairness within a single ranking, which can then be aggregated to compute system fairness; others directly assess the fairness of a sequence or distribution of rankings.
%Not all have been developed in the context of information access --- some are intended for single-shot rankings such as college rankings or university admissions, but can be adapted to information access settings.
%These metrics mostly focus on two IR relevant objectives of group fairness: \emph{demographic or statistical parity} ensures comparable outcomes across groups and \emph{equality of opportunities} ensures equal treatment based on merit or utility irrespective of the group membership.
%The key difference between these metrics lies in how they relate fairness to relevance: some only measure the equality of exposure, and must be integrated with other metrics to account for relevance, while others incorporate relevance into the metric itself to measure exposure commensurate with relevance; which is desired depends on the specific application and fairness goals.
%Table~\ref{tbl:summary} summarizes the metrics we consider.
%Metric names are chosen based on their functionality, purpose, and comparability within our synthesis; in some cases, we use the original name, but in others we assign a new name name either because the original paper used a general name (e.g. ``unfairness'') or the original name does not make for clear exposition in comparison to other metrics and relevance. We describe these metrics in detail later in this section.
\subsection{Problem Formulation}
\begin{table}[tb]
\centering
 
 \caption{Summary of notation.}
 \begin{tabular}{cl}
 \toprule
  $\doc \in \allitems$ & document or item \\
  $\user \in \allusers$ & request (query or user) \\
  $\ranking$ & ranked list of $N$ documents from $\allitems$ \\
  $\ranking^{-1}(i)$ & the document in position $i$ of list $L$ \\
  $\rankingPosition$ & rank of document $d$ in $L$ \\
  $\ranking_{\le k}$ & prefix of $\ranking$ of length $k$ \\
  $\relevance$ & relevance of $d$ to $q$ \\
  $\group$ & number of groups \\
  $\alignmentvec$ & group alignment vector \\
  $\alignmentmat$ & group alignment matrix for documents in $\ranking$ \\
  $\Protected$ & set of documents in protected group in $L$ \\
  $\nonProtected$ & set of documents non-protected group in $L$ \\
  $\populationEstimator$ & target group distribution \\
  %$\probabiltyDist(\ranking)$ & probability distribution of groups in $\ranking$ \\
  $\attentionmat$ & attention vector for documents in $\ranking$ \\
  $\attentionvec$ & position weight of $\doc$ in $\ranking$ \\
  $\groupexposure$ & the exposure of groups in $\ranking$ ($\alignmentmat^T \attentionmat$) \\
%   $\Upsilon_S$ & is the utility of the ranked list $S$
%   $\epsilon$,$\hat{\epsilon}$ is the expected system exposure and expected target exposure. 
%  %This can be computed by weighting the ranking positions with different models i.e. \textit{logarithmic loss} or \textit{Geometric distribution} or different user browsing models.
%   $a(d)$ is the estimated attention received by item $d$ at ranking position $i$. 
%   In case of binary and explicitly known membership, $G^+$ can be the set of items in protected group and $G^-$ refers to the set of remaining items where $|G^+|=\sum_{i\in L}G_i$.
\bottomrule
 \end{tabular}
 \label{tab:notation}
\end{table}
We consider an IR system that retrieves a ranked list $\ranking$ of $n$ documents $\docs{1}, \docs{2}, \dots, \docs{n}$ $\in \allitems$ in response to requests (query in a search system or a user in recommender system) $\users{1}, \users{2}, \dots, \users{n} \in \allusers$ (notation summarized in table~\ref{tab:notation}).
%$d_1, d_2, \dots, d_n \in I$ for requests $q_1, q_2, \dots, q_m \in Q$ (a request is a user profile in a recommender system or a query for a search application).
% A response comes in the form of a ranked list $\ranking$ of $N$ documents from $\allitems$.
%$\ranking(d)$ is the 1-based rank of document $d$ in $\ranking$, and $L^{-1}(i)$ is the document at position $i$.
%Some metrics consider prefixes of $\ranking$; $\ranking_{\le k}$ denotes the prefix of $\ranking$.
Documents may have an associated request-specific relevance score $y(d|q)$, and the system may estimate this by a predictor $\hat{y}(d|q)$.
Providers are associated with one (or more) of $\group$ groups. We represent this by giving each document an alignment vector $\alignmentvec \in [0,1]^g$ (s.t. $\|\alignmentvec\|_1=1$) indicating its group association; generalizing from a categorical variable to a vector allows soft association (mixed or partial membership) or uncertainty about membership \citep{sapiezynski2019quantifying}. We generalize $\group$ to a list function, with $\alignmentmat$ denoting an $n \times \group$ alignment matrix whose rows correspond to the documents of $\ranking$ and columns are groups.
In the case of definitively-known membership in a binomial pair of groups, $\Protected$ denotes the set of documents in $\ranking$ in the ``protected'' group and $\nonProtected$ the remaining documents.
%The system exposes documents to the user by placing them in the ranked list, and users may or may not give them attention; With available data in an offline experiment, what we can measure is the fairness of exposure, even though it may be called attention in some prior work. 
 %Group membership is often defined by \textit{sensitive attributes} such as race, gender, or ethnicity.

Our goal is to measure \emph{exposure} (sometimes called \emph{attention}) each document, content provider, or group receives, and assess the fairness of this distribution to ensure demographic or statistical parity (ensures comparable outcomes across groups or \emph{equality of opportunities} (ensures equal treatment based on merit or utility irrespective of the group membership).
Accounting for the decreasing attention users are likely to pay to documents at deeper rank positions (\emph{position bias}) requires a browsing model; some metrics build this implicitly into their structure, while others explicitly model it as a \emph{position weight vector} $\attentionmat$ for $\ranking$. Table~\ref{tab:weighting} describes the various weighting schemes used by the metrics we survey.
The resulting exposure is then sometimes compared with a \emph{target distribution} $\populationEstimator$ that represents across groups. 
There are several ways of computing $\populationEstimator$, including strict group equality, an estimate of the distribution of actual or potential content providers, or the distribution among providers of relevant documents.

\begin{table*}[tb]
    \centering
    \caption{Weighting models for computing $\attentionvec$.}
    \label{tab:weighting}
    \begin{tabular}{lcl}
    Model & Formula & Parameters \\
    \toprule
        Geometric & $\gamma (1-\gamma)^{\rankingPosition - 1}$ &
        Stopping probability $\gamma$ \\
        Logarithmic & $1 /  \operatorname{log}_2 \operatorname{max}\{\rankingPosition, 2\}$ &
        --- \\
        RBP & $\gamma^{L(d)}$ &
        Continuation probability (patience) $\gamma$\\
        Cascade &
        $\gamma^{\rankingPosition - 1}\prod_{j \in [0,\rankingPosition)} \left[1-\phi\left(y(\ranking^{-1}(j)|y)\right)\right]$ &
        Patience $\gamma$, stopping probability function $\phi$ \\
    \bottomrule
    \end{tabular}
\end{table*}

\begin{table}[]
\caption{Distance functions for comparing distributions.}
\label{tab:dist}
\begin{minipage}{\columnwidth}
\centering
\begin{tabular}{ccc}
Distance Function  & $\populationEstimator$\footnote{Binomial $\populationEstimator$ is a scalar probability of the protected group.} & Formula                                                                                                                                                                                                                      \\ \hline
$\Delta_{\mathrm{ND}}(L,\populationEstimator)$  & Binomial      & $\frac{|\Protected|}{N} - \populationEstimator$                                                                                                                                                                            \\ 
$\Delta_{\mathrm{RD}}(\ranking,\populationEstimator)$  & Binomial      & $\frac{|\Protected|}{|\nonProtected|}-\frac{\populationEstimator}{1 - \populationEstimator}$                                                                                                                                                        \\ 
$\Delta_{\mathrm{KL}}(L, \populationEstimator)$ & Multinomial       & 
$D_{\mathrm{KL}}(\populationEstimator(\ranking) \Vert \populationEstimator)$\footnote{K-L divergence; $\populationEstimator(\ranking)$ is the probability distribution of groups in $\ranking$.} \\ 
%$\Delta_{\mathrm{AD}}(\groupexposure, \populationEstimator)$              & Binomial       & 
%$\lvert\frac{\lvert \Protected \rvert}{N} - \populationEstimator \rvert$
%$\epsilon(L) = G(L)^{\mathrm{T}} a(L)$                                                                                                                                                                                       \\ \hline
\end{tabular}
\end{minipage}
\end{table}

\subsection{Statistical Parity in Single Rankings}
%%The fallowing two works we discuss measure fair ranking by looking at the \emph{composition} of the rankings. They produce a fairness metric for an individual ranking, which can then be aggregated over multiple rankings to measure a system's overall fairness. 
We begin with metrics that assess the fairness of a single ranking and only measure exposure equity without considering relevance (that is, they target \textit{statistical parity}). These metrics can be aggregated over the rankings produced by a system, e.g. by taking the mean, to produce an overall system fairness score. 

The simplest way to measure the fairness of a single ranking is to measure the proportion of items in each group~\citep{ekstrand:bag}, but this does not account for position bias.
\citet{yang2017measuring} propose a family of statistical parity measures that incorporate position bias by averaging parity over successive prefixes of the ranking; we therefore call this the \textit{prefix fairness} family ($\prefd$).
These metrics are optimized when the representation in each prefix matches the target $\populationEstimator$ as closely as possible, as measured by a distance function $\Delta$; \citeauthor{yang2017measuring} used the full ranking's composition as $\populationEstimator$, 
%group-fairness aware ranking based on the assumption that items at the high ranking positions receive advantages over the low-ranked items. With that intuition they present three related measures of \emph{statistical parity} between groups in a single ranking. To prioritize parity at the top of the list, these measures average the parity over successive prefixes of the ranking; we call them the \textit{prefix fairness} family.
and instantiate $\prefd$ with distance functions ND, RD, and KL (from Table~\ref{tab:dist}) to yield different members of the family.
The metric is defined as
\begin{align}
    \label{eq:pref}
    %\mathrm{\hat{\mathbf p}} = \frac{\Delta(L_{<=k}, L)}{log_2i}\\
    \prefd(\ranking) = \frac{1}{Z}\sum_{i=10,20,30,...}^{N}\frac{\Delta(\prefix{i}, \probabiltyDist)}{log_2i}
\end{align}
\noindent
where normalizing scalar $Z = \max_{\ranking'} {\prefd'(\ranking', \probabiltyDist)}$ (taken over all $L'$ with the same length and group composition as $\ranking$, where $\prefd'$ is the prefix fairness function without the normalizer), scaling $\prefd$ to the range $[0,1]$ where $1$ represents maximum unfairness.
$\Delta_{\mathrm{KL}}$ has the advantage of allowing multinomial protected attributes and soft group association.
\emph{$\prefd$} does not work when $\nonProtected = \emptyset$, and $\Delta_{\mathrm{RD}}$ does not work when $\nonProtected$ is small. The normalizing constant is troublesome to compute with incomplete group membership data.

\citet{zehlike2017fa} propose a similarly-motivated group fairness constraint for single list and binomial groups: $\ranking$ satisfies the FAIR constraint if for every prefix $\prefix{k}$ with $1\le k \le N$, the protected group is not statistically significantly under-represented.  Unlike $\prefd$, FAIR does not penalize over-representing the protected group.
%that is, the probability that $k$ Bernoulli trials with probability $\populationEstimator$ will have $m=\|\groups^+(\prefix{k})\|$ or fewer successes with probability at least $\alpha$.
%measures group fairness of a single list by comparing the proportion of protected candidates in \textit{every} prefix of the top-n ranking to a given minimum proportion $\probabiltyDist$. They presented their construct as a constraint: given a minimum proportion $\probabiltyDist$, a ranked list $\ranking$ satisfies the FA*IR constraint if for every prefix $\prefix{k}$ with $1\le k \le N$, the protected group is not statistically significantly under-represented; that is, the probability that $k$ Bernoulli trials with probability $\probabiltyDist$ will have $m=\|\groups^+(\prefix{k})\|$ or fewer successes with probability at least $\alpha$.
We convert this constraint into a metric by taking the average of the binomial probabilities:
\begin{align}
    \label{eq:FAIR}
    \mathrm{FAIR}(L) & = \frac{1}{N} \sum_{k=1}^N P_{\mathrm{Binomial}}\left(m \le \lvert \groups^+(\prefix{k})\rvert \cond \probabiltyDist, k\right) \\
    \nonumber
    &  = \frac{1}{N} \sum_{k=1}^N \sum_{j=1}^{|\groups^+(\prefix{k})|} {k \choose j} (\probabiltyDist)^{j} (1-\probabiltyDist)^{k-j}
\end{align}
FAIR limits applicability into binomial distribution and fixed group association. 

\citet{sapiezynski2019quantifying} provide a more general metric for single-list fairness by using an explicit (and configurable) position weight model $\attentionmat$ instead of embedding the browsing model in the metric structure.
% They generalize from the distribution of the entire ranked list to a $\populationEstimator$ %a distribution over groups that is considered optimally fair.
% $\populationEstimator$ is computed from the set of all relevant items, equivalent to the use of the whole ranking in Eq. \ref{eq:pref}; the set of all content producers; the population at large; or other means of assessing the target group representation.
%Table~\ref{tab:dist} presents the distance function for measuring $\AWRF$.
Given an alignment matrix and suitably normalized position weight vector, $\groupexposure = \alignmentmat^{\mathrm{T}} \attentionmat$ is a distribution that represents the cumulative exposure of the various groups in $\ranking$.
The resulting unfairness metric, which we call \textit{Attention-Weighted Rank Fairness} ($\AWRF$), is the difference between this exposure distribution and the population estimator:
\begin{equation}
\label{eq:AWRF}
    \AWRF(\ranking) = \Delta(\groupexposure, \populationEstimator) 
\end{equation}
$\AWRF$ allows soft association and multinomial protected attributes.
The distance function in Table~\ref{tab:dist} depends on application context; for assessing a particular protected class representation, difference in probability is suitable distance.
%K-L divergence can be used for multiple groups.
%\noindent \fbox{
%\parbox{\textwidth}{
%		\textit{Key Takeaways:} \emph{$\prefd$}, \emph{FAIR}, and \emph{$\AWRF$} measure fairness in single-list ranking without considering relevance information. \emph{FAIR} differs from $\prefd$ by considering every prefix instead of prioritizing top-orders only. \emph{$\AWRF$} differs from other two by taking position weight into account.}}

\subsection{Statistical Parity in Multiple Rankings}
In many cases, fair exposure cannot be achieved in a single ranking, because the attention paid to rank positions often decreases more steeply than utility \citep{biega2018equity, diaz2020evaluating}.
One way to address this is to measure fairness over sequences or distributions of rankings, so providers have comparable opportunity to be exposed in response to at least some requests.
This approach can be modeled as a request-dependent distribution (or \textit{policy}) $\stochasticRanking(\ranking|\user)$ over rankings \citep{singh2018fairness, diaz2020evaluating}.
We extend this to also model the arrival of requests as a distribution $\rho(\user)$, so a sequence of rankings $\ranking_1,\ranking_2,\dots,\ranking_{\tilde n}$ \citep{biega2018equity} is a series of draws from the distribution $\rho(\user)\stochasticRanking(\ranking|\user)$.
The group exposure within a single ranking from Eq. \ref{eq:AWRF}, $\groupexposure = \alignmentmat^T \attentionmat$, is the fundamental building block of these metrics, along with its expected value:
\begin{align*}
\exposure(\user) & = \expectation{\stochasticRanking}{\groupexposure} = \sum_\ranking \stochasticRanking(\ranking \mid \user) \groupexposure \\
\exposure_\stochasticRanking & = \expectation{\stochasticRanking\rho}{\groupexposure} = \sum_\user \rho(\user) \exposure(\user)
%\exposure & = \mathrm{E}_{\stochasticRanking\rho}[\epsilon(\ranking)] = \sum_\user \rho(\user) \sum_\ranking \stochasticRanking(\ranking \mid \user) \groupexposure
\end{align*}
% \subsection{Sequence Metrics Statistical Parity}

\citet{singh2018fairness} and \citet{diaz2020evaluating} each propose metrics for measuring statistical parity over ranking policies. Neither metric incorporates a target distribution; they are optimal when all groups are equally exposed. 
Demographic parity \citep[DP,][]{singh2018fairness} measures the difference in exposure between two groups:\footnote{The original paper presented a constraint, not a metric, for demographic parity; we have implemented it as a ratio to be consistent with the other metrics.}
\begin{equation}
    \label{eq:dpdiff}
    \mathrm{DP} = \exposure_\stochasticRanking(\progroup) / \exposure_\stochasticRanking(\nonprogroup)
\end{equation}
Expected exposure disparity \citep[EED, ][]{diaz2020evaluating} ensures well-distributed exposure by measuring the inequality in exposure distribution across groups with the $L_2$ norm:
\begin{equation}
    \mathrm{EED} = \|\exposure_\stochasticRanking\|_2^2
\end{equation}
%and 
%Using the metrics that do not account for relevance are best suited for measuring the relative fairness of rankings optimized for utility, particularly when there are large relevant sets but using them in isolation for evaluation or optimization may reduce ranking quality.
\subsection{Equal Opportunity in Multiple Rankings}
So far, none of the metrics we have discussed account for the utility of the ranked results --- rankings do well by exposing providers regardless of the utility of their items. %The exposure family of metrics operates over distributions or sequences of rankings, and directly incorporate relevance in various ways. 
The intuition behind incorporating utility, articulated independently by \citet{singh2018fairness} and \citet{biega2018equity}, is that exposure should be proportional to relevance: if an item or a group contributes 10\% of the relevance to a request (user and/or query), it should receive approximately 10\% of the exposure.
This is a ranked analog of the \textit{equality of opportunity} construct from fair classification \citep{hardt2016equality}: outcome is conditionally independent of group given utility.
To measure deviation from this goal, \citet{singh2018fairness} propose two metrics.  The \emph{exposed utility ratio} (EUR)\footnote{\citet{singh2018fairness} used the terms ``disparate treatment ratio'' and ``disparate impact ratio'' for EUR and RUR, respectively; this terminology is not consistent with the use of these terms in the broader algorithmic fairness literature; as we understand it, exposure the system gives to providers is an impact, not a treatment. We have changed the names to reduce confusion going forward.}, measures violation from the goal that each group's exposure is proportional to its contributed utility (measured by $\Upsilon(\groups) = \mathrm{E}_\rho[\frac{1}{\group} \sum_{i \in \group} \relevance]$):
\begin{equation}
\label{eq:eur}
    \mathrm{EUR}=\frac{\exposure_\stochasticRanking(\progroup)/\Upsilon(\progroup)}{\exposure_\stochasticRanking(\nonprogroup)/\Upsilon(\nonprogroup)}
\end{equation}
The \emph{realized utility ratio} (RUR) incorporates utility into both numerators and denominators by measuring whether the \textit{discounted} utility contributed by each group ($\Gamma(\groups) = \sum_{\doc\in \groups} \mathrm{E}_{\stochasticRanking\rho}[\attentionvec \relevance]$) is proportional to its total utility:
\begin{equation}
\label{eq:rur}
    \mathrm{RUR} = \frac{\Gamma(\progroup)/\Upsilon(\progroup)}{\Gamma(\nonprogroup)/\Upsilon(\nonprogroup)}
\end{equation}
EUR and RUR do not allow multinomial protected groups and soft associations. 
%RUR is highly vulnerable to the relevance sparsity problem.

\citet{biega2018equity} present the \emph{amortized attention} construct to measure exposure over the sequence of rankings.
%As discussed above, we can treat the sequence as a sequence of draws from the distribution of requests and rankings and replace their sums with expectations (as summing over the sequence is equivalent to expectation multiplied by sequence length).
This compares rank exposure with expected utility $\hat\Upsilon$ (computed with system-predicted utility $\predictedrelevance$) instead of ground truth relevance assessments $\relevance$), measuring whether the system allocates exposure proportional to the utility it estimates items to have.
%Given the expected relevance $\hat\Upsilon$ (computed as above, except with estimated relevance), they set the goal that $\frac{\exposure_\stochasticRanking(\groups^1)}{\hat\Upsilon(\groups^1)} = \frac{\exposure_\stochasticRanking(\groups^2)}{\hat\Upsilon(\groups^2)}$ for all pairs of groups $\groups^1, \groups^2$; 
Deviations from this goal are measured by taking the $L_1$ norm of the group exposure-utility differences, yielding the \textit{Inequity of Amortized Attention} (IAA) metric:
\begin{equation}
    \mathrm{IAA} = \|\exposure - \Upsilon\|_1
\end{equation}
% \emph{IAA} allows multinomial protected groups. 
%They use a geometric decay for position weights.
%The original focus of the amortized attention framework is individual fairness, where each item (or author) is exposed proportionally to their relevance, but group fairness is a groupwise aggregate of individual fairness under this definition.

\citet{diaz2020evaluating} build on this to integrate relevance in a different way.
Rather than relate exposure directly to relevance, they use relevance to derive \emph{target exposure} based on an ideal policy $\tau$ that assigns equal probability to all rankings that place items in non-decreasing order of relevance and $0$ probability to all other rankings.
The target exposure $\exposure^*$ is the expected exposure under the ideal policy ($\exposure^* = \mathrm{E}_\tau\rho[\groupexposure]$).
They take the squared Euclidean distance between system expected exposure and target exposure, yielding the \emph{Expected Exposure Loss}:
\begin{align}
    \mathrm{EEL} & = \|\exposure_\stochasticRanking - \exposure^*\|_2^2 \\
    \label{eq:eel-decomp}
    & = \|\epsilon_\stochasticRanking\|_2^2 - 2\exposure_\stochasticRanking^\intercal\exposure^* + \|\exposure^*\|_2^2
\end{align}
The decomposition in Eq. \ref{eq:eel-decomp} yields \emph{expected exposure relevance} $\mathrm{EER} = 2\exposure_{\stochasticRanking}^{\mathrm{T}}\exposure^*$ (measuring the alignment of exposure and relevance, with higher values representing better distributions) along with EED.
%They propose two models for the position weights, a cascade model based on expected reciprocal rank \cite{chapelle2009expected} and a geometric model from rank-biased precision \cite{moffat2008rank}.

Neither IAA nor the EE metrics distinguish between group over- or under-exposure; for both, $0$ is perfectly fair and larger values are unfair, with no preferential treatment given to a protected group.
The common thread between these metrics, articulated by \citet{diaz2020evaluating}, is that for a fixed information need, differences in exposure between items with the same relevance grade results in unfair outcomes.
The only way to address this inequity in practice is by varying the rankings returned by the system, as with a stochastic policy.
\subsection{Pairwise Metrics}
\citet{beutel2019fairness} and \citet{narasimhan2020pairwise} take an entirely different approach by defining fairness objectives over pairwise orderings instead of entire rankings.
Pairwise fairness is then defined in terms of the pairwise accuracy for ranking relevant items in different groups:
\begin{align}
    % A_{G_1>G_2} = P\left(L^{-1}(d_1) > L^{-1}(d_2) \mid y(d_1\mid u) > y(d_2\mid u), d_1 \in G_1, d_2 \in G_2\right)
    A_{\groups^1>\groups^2} & = P\left(\docs{1} \succ_\ranking \docs{2} \cond \docs{1} \succ_{y \mid \user} \docs{2}, \docs{1} \in \groups^1, \docs{2} \in \groups^2\right)
\end{align}
%where
\begin{align*}
    \docs{1} \succ_\ranking \docs{2} & = \ranking(\docs{1}) < \ranking(\docs{2})
    & \text{$\docs{1}$ ranks above $\docs{2}$}
    \\
    \docs{1} \succ_{y \mid \user} \docs{2} & = y(\docs{1} \mid \user) > y(\docs{2} \mid \user)
    & \text{$\docs{1}$ more relevant than $\docs{2}$}
\end{align*}
A ranking satisfies \emph{pairwise equal opportunity}\cite{narasimhan2020pairwise} if pairs of documents are equally likely to be ranked consistently with their relevance regardless of the group membership of the items in the pair.
This can be measured by the group's pairwise accuracy with respect to all items ($A_{G^i>:})$, its \textit{inter-group} accuracy ($A_{\groups^1>\groups^2}$), or its \textit{intra-group} accuracy ($A_{\groups^1>\groups^1}$).
Given protected and unprotected groups, we can define a fairness metric as the difference in pairwise accuracy:
\begin{align*}
    %\mathrm{PairAcc} & = A_{\nonprogroup>:} - A_{\progroup>:} \\
    \mathrm{IntraAcc} & = A_{\nonprogroup>\nonprogroup} - A_{\progroup>\progroup} \\
    \mathrm{InterAcc} & = A_{\nonprogroup>\progroup} - A_{\progroup>\nonprogroup} \\
\end{align*}
%\noindent \fbox{
%\parbox{\textwidth}{
%		\textit{Key Takeaways:} \emph{PAIR} considers single ranking and relevance of items.
%		}}
%that may be easier to apply with missing relevance and/or group membership data.
%It requires further adaption to fit within our current experimental setting, which we leave for future work.
\subsection{Assessing Metric Design}
Rendering metrics in a common notation elicits that the metrics are quite similar in their basic concepts.
The fundamental construct --- weighted exposure --- is the same across most metrics (pairwise fairness being an exception), and they differ primarily in how they relate exposure to relevance and how they aggregate and compare exposure distributions.
The following questions help identify more precisely what their salient differences are and how those may relate to particular IR applications and experimental settings.
% We identified some questions that will help us to distinguish the metrics from each other when it comes to implement in real IR setup.
%\begin{itemize}
 %   \item Q1: Does it incorporate relevance?
  %  \item Q2: What is target?
  %  \item Q3: How does it compare the system to target?
%\end{itemize}

\textbf{Does the metric incorporate relevance?} EEL, EER, EUR, RUR, IAA, and PAIR directly incorporate relevance into the metric; others strictly measure statistical parity. It is desired depending on the precise task and evaluation goal. Statistical parity metrics are useful for measuring relative fairness of rankings already optimized for utility, particularly when there is no relevance information available or large relevant sets available. They can also be used to detect discrepancies that may indicate unfairness in relevance data. However, using such metrics in isolation for evaluation or optimization may reduce ranking quality.
%Using metrics that use relevance opens up further concerns about handling missing relevance information and relevance source.

\textbf{How does it handle missing data?} Real-world data sets are often incomplete, missing relevance and/or group labels for many documents. Metrics that are less vulnerable towards that problem will be easier to apply in such cases. Missing relevance data affects EUR, RUR, EER, and EEL like it does classical IR evaluation metrics such as nDCG; the straightforward but biased approach is to treat items with unknown relevance as irrelevant ($y=0$). IAA's use of system-estimated relevance allows it to sidestep missing relevance problem.

Missing group labels require different handling. For many metrics we can include unlabeled items when computing attention weights but exclude them from further analysis, or treat ``unknown'' as an additional group identity.
Unknown data is a more significant problem for the $\prefd$ family because it treats a list with fewer than 10 known-group items as maximally fair, and the straightforward way of computing $Z$ --- make the ranking maximally unfair by putting all protected items last --- does not work in the face of missing data. 

\textbf{How does it respond to edge cases?} 
Realistic IR experiments bring a number of important edge cases, such as groups with no items relevant to or retrieved for a request.
Ratio-based metrics and distance functions are particularly vulnerable to these problems; EUR metric and $\Delta_{\mathrm{RD}}$ distance function, for example, approach infinity as the number of dominant-group protected items retrieved goes to zero.
RUR is even more brittle, as it requires nonzero relevance from retrieved dominant-group items to avoid infinity, and both it and EUR can be infinite or undefined if the set of relevant items from either group is zero.

\textbf{What is the target?} $\prefd$, FAIR, $\AWRF$, EEL, and EER provide flexibility in determining how the (un)fairness of exposure is ultimately assessed through selection of the target distribution, while targets are implicitly baked in to the structure of others. This configurability is useful because it allows the metric to be adapted to the fairness requirements of a particular task, although it can impair comparability between experiments.
% For example, IAA targets proportionality to the system's predicted relevance; if relevance estimates are unfair (e.g. under- or over-estimating relevance for certain groups), IAA's assessment will be inaccurate.  If a model of this bias is available, it can be incorporated into the target distribution. and if the estimated relevance are biased in an unfair way, the metric's assessment of fairness will be impacted. Most of these metrics can be implemented with other relevance judgements.
%although EEL and EER are not applicable to estimated relevance because the system will generally rank in order of estimated relevance, thus always obtaining ideal rankings.

\textbf{How does the metric compare the system with target?}
Some metrics ($\AWRF$ and $\prefd$) use an explicit distance function to compare distributions, while others use ratios of specific proportions or norms of differences in distributions. Norms and selected distance functions (such as $\Delta_{\mathrm{KL}}$) can accommodate soft association, while ratios and distance functions based on binomial probabilities require definitive membership in binomial groups.
They can be adapted to some multi-group situations if only one group's exposure needs to be considered.
% The norm of a difference, as in IAA and EEL, or a difference in distributions as in $\AWRF$, can be used to measure fairness with respect to soft association, unlike ratios or binomial probabilities.

\textbf{What user model does it use?} 
Most metrics allow different position weighting strategies to be selected, both in its structure and its parameters.
This configurability allows the metric to be adapted to a specific application but introduces potential sensitivity to choices of weight functions and parameter values.
$\prefd$, FAIR, and PAIR are not configurable, as position weighting is built-in (as in $\prefd$ and FAIR) or unavailable (in PAIR).
\section{Empirical Comparison}
To complement our comparison, we implement the metrics to measure the fairness of systems from prior experiments in search and recommendation.
%
%\begin{itemize}
%    \item FairTREC \citep{trec-fair-ranking-2019, biega2020overview}, a track at TREC 2019 and 2020 in which submitted systems were to respond to scholarly search queries with rankings that are fair with respect to economic development level of the authors' countries of operation.
%    \item The GoodReads experiments by \citet{ekstrand:bag}, measuring the fairness of book recommendations with respect to author gender.
%\end{itemize}
%We further carry out a sensitivity analysis to understand how experimental outcomes change in response to design decisions and parameter values in the metrics.

\subsection{Data and Tasks}
\begin{table}[tb]
    \caption{Summary of experiment data.}
    \label{tab:data}
    \begin{minipage}{\textwidth}
    \centering
   
\begin{tabular}{l@{\hskip 0.5in}c@{\hskip 0.5in}c@{\hskip 1em}c@{\hskip 1em}c}
& & \multicolumn{3}{c}{Fair TREC} \\
& GoodReads & 2019 Rerank & 2020 Rerank & 2020 Ad-hoc  \\
\hline
Systems & 4 & 14 & 23 & 5 \\
Requests & 5000 & 500 & 195 & 189 \\
Items         & 23,60,655          & 2052               & 2112      &  2112                  \\
$|\progroup|$  & 1,90,711           & 312                 & 294       & 294                   \\
$|\nonprogroup|$ & 21,17,451          & 1906                & 1632    & 1632                     \\ \hline

    %\bottomrule
    \end{tabular}
    \end{minipage}
\end{table}
Table~\ref{tab:data} summarizes our experimental data.
For search experiments, we used submitted runs and evaluations from the \textit{TREC Fair Ranking Track} 2019 \citep{trec-fair-ranking-2019} and 2020 \citep{biega2020overview}.
These runs covered three tasks across two years, all on scholarly search: re-ranking tasks from 2019 and 2020 and an ad-hoc retrieval task from 2020.
Each document has a soft association with the economic development level of its author(s), and we consider each submitted run as an individual system and used the given sequences of rankings for each system.
%There were 14 submissions in total, each we consider as individual algorithm. 

For the recommendation task, we re-used the experiment and code from \citet{ekstrand:bag} to measure 4 algorithms from the LensKit software \citep{ekstrand2020lenskit} on data from GoodReads \citep{wan2018item}, modifying the experiment to use 5 interactions per user as test data instead of the original 1. Group membership is binary, with female authors as the protected group\footnote{Binary gender membership is a limitation of the original study; see \citet{ekstrand:bag} for details.}; membership is unknown for some items.
%We use LensKit for Python \citep{ekstrand2020lenskit} to generate recommendations for users in the \emph{GoodReads} book data \citep{wan2018item}.
%We train the recommendations on implicit feedback data from GoodReads, with a positive user-book interaction if the user have added the book to a shelf.
We sampled 5000 users, each of which had at least 10 book interactions, for our experiment,
holding out 5 interactions per user as test data for assessing relevance in the resulting recommendations.
%Author gender information of the books is extracted from Virtual Internet Authority File (VIAF)\footnote{http://viaf.org/viaf/data/} as described by \citet{ekstrand:bag}.
We measured the fairness of the recommendations produced for these users by four collaborative filtering (CF) algorithms: user-based CF (UU, \cite{10.1145/312624.312682}), item based CF (I-I, \cite{deshpande2004item}), matrix factorization (WRLS, \citep{Takacs2011-ix}), and Bayesian Personalized Ranking (BPR, \cite{rendle2012bpr}), using algorithms and hyper-parameter tunings from \citet{ekstrand:bag}. 
%We generated 100 recommendations for each user.

\subsection{Metric Implementations}

We implemented metrics from Section \ref{sec:metrics} in Python to measure the fairness of the runs from the experiments we consider.
%For each metric's configurable parameters, we used the settings from the original paper; determining how to configure the metrics for specific applications is future work. 

While implementing $\prefd$ metric, we had difficulties because this metric suffers from missing data and numerous edge case breakdown. We exclude $\prefd$ for further analysis.

We did need to make some further decisions and adjustments to the remaining metrics in our experiment:
\begin{itemize}
\item For $\AWRF$, we follow the original implementation and use $\Delta_{\mathrm{ND}}$ to compare the representation of the protected group with the target distribution.
We applied $\AWRF$ with two target distributions: $\AWRF$ computes $\populationEstimator$ from the distribution of providers in the full data set, while \textsf{AWRF\_equal} targets equal representation of protected and unprotected groups. 
\item For IAA and the EE* metrics on the recommendation experiment, we treat unknown gender as a third author group.
\item 
PAIR do not depend only on the top-$N$ list --- they are functions of the system's overall ranking between items. Therefore, instead of computing them from ranked output we directly measured the recommendation model's scores for a sample of items. For each test item, we sampled 10,000 items not rated by the target user as negative examples, and used these to estimate the probability of correct orderings. This proved relatively efficient for our experiment size.
We could not test these metrics on FairTREC because we do not have access to full rankings or the systems' relevance scores.
\end{itemize}
% We were not able to obtain useful results using pairwise accuracy without a sufficient number of rated items.  In order to solve this issue, we used a large number of recommendation lists to obtain an amount that merits usage of the metric.  However, this came with the cost of long computation time for large recommendation lists.
%rather than being able to lower the number of recommendations to a healthy amount for sane analysis.
\subsection{Empirical Results}
\begin{figure*}[th!]
    \centering
    \includegraphics[width=\textwidth]{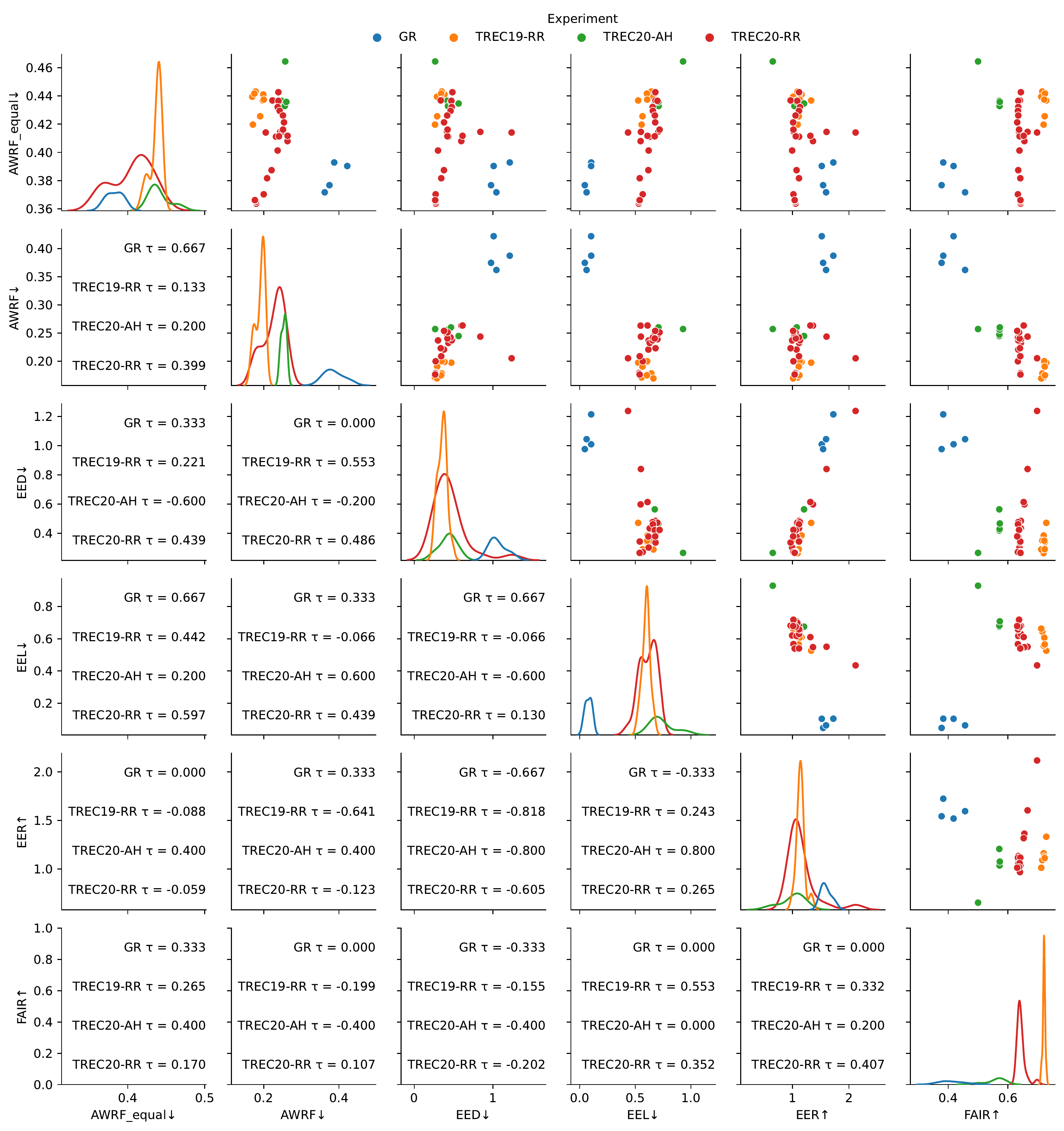}
    \caption{Metric results and correlations.  Arrows indicate the direction of increasing fairness.  Correlations computed with Kendall's $\tau$-c within each experiment, ordering systems according to each metric's directionality.}
    \label{fig:default}
\end{figure*}
%We now present the results of our experiment, using the metrics in their default configurations (parameter settings from their original papers). 
Our goal is to observe whether the metrics agree or disagree, and if this agreement is consistent across experiments. Figure~\ref{fig:default} shows the metric results from our experiments along with their Kendall $tau$ correlations. We do not show results from metrics that only worked on one experiment, but the metrics do not show clear consensus across datasets; there are substantial differences in their system orderings, and metrics that agree in one experiment don't often don't agree on others.
The most consistently agreeing pair is EEL and $\AWRF$ with an equal-exposure target (positive correlation in all experiments, and comparatively high correlation in three of them).
\section{Conclusion, Recommendations, and Future Directions}

Many metrics have been proposed for measuring the fairness of ranked outputs.  Through analytically and empirically comparing fair ranking metrics, we have identified several of their key commonalities and differences, assumptions and crucial design decisions, and have presented them in a form amenable to implementation in IR experiments.  Our empirical results highlight the need for further research to identify the implications of these metrics' disagreements, as well as the impact of specific design decisions and configurations, but our analytical comparison (Section~\ref{sec:metrics}) and implementation experience allow us to make some preliminary recommendations for selecting a metric:

\textbf{Single Rankings} 
$\AWRF$ seems the most generally useful, as it supports multinomial protected attributes with soft association, and is adaptable to multiple attention models, target distributions, and difference functions depending on the application.
We are not yet able to make concrete recommendations for the choice of difference function.

\textbf{Demographic Parity in Sequences} Again, due to support for multinomial groups and soft association, along with edge-case problems in ratio-based metrics, \textbf{EED} looks like the best current choice for this case.

\textbf{Equal Opportunity in Sequences}
We currently recommend using \textbf{EEL} and \textbf{EER} because they allow multinomial groups with soft association, and selectable target distribution.
If sparse relevance judgments are a problem, IAA is a good choice and can be applied with multinomial groups and soft associations.
% This paper presents a comparative analysis of several fair ranking metrics. We discuss the metric formulations and implications in an integrated framework and presented an empirical comparison with a common data set and fairness goal. 
% Defining the metrics under uniform notation gave us in-depth knowledge about the fairness goal, assumptions, and design decisions. While doing that, we noticed that some metrics are surprisingly similar in their underlying concept of fairness. 
% The metric implementations under the same experimental setup elicits the applicability of the metrics in a real dataset. We recommend some metrics for given task based on their applicability.
%We learnt that the the differences in fairness goal and the meaning of direction in these metrics makes it difficult to interpret and directly compare results. However, we identified crucial factors such as group size, ranked list size, item relevance information, and group membership can have substantial effect on implementing the metrics. 
%Our study presents the applicability of the fair ranking metrics in real data focusing on the strengths and limitations while implementing them. 
%We started this project with the goal of identify the requirements to implement the fair ranking metrics in actual IR frameworks.
%Identify the similarities and empirical differences among the metrics to ease the metric selection process.
  %\item Identify the influential factors in the metric design decision making process.

There is significant further work needed to advance the robust measurement of ranking fairness.
Data sparsity and ambiguity are a significant problem; there has been work on allowing missing \cite{kirnap2021estimation}, ambiguous, or multiple group associations \cite{ghosh2021fair} that needs to be incorporated into flexible metrics.
We also need further study on the implications of specific design and parameter choices; our empirical results have shown notable disagreement between metrics, but we do not yet know which of the various design decisions is responsible for these differences, or how metric values change in response to particular aspects of the data set or system rankings. 
% Simulation study on these metrics, will help to understand the impact of factors like, relevant set size, soft association, missing relevance information etc.
%Furthermore, considering alternative ranking models may introduce more complexity in measuring fair ranking.
%Significant progress has been made in the last 2--3 years on measuring the fairness of rankings, but more work is needed in order to understand how best to design and apply these metrics. We hope this paper provides researchers with useful guidance in navigating and building on this space.
%The sensitivity analysis enables us to better understand the metric sensitivity and dependency on various design aspects. The observation from the sensitivity analysis results showed that despite having similar fairness goals, these metrics can differ in their sensitivity towards external factors. 

We believe our synthesis and comparison will lay a valuable foundation for this vital and ongoing work.

\begin{acks}
This material is based upon work supported by the National Science Foundation under Grant No. IIS 17-51278.
\end{acks}

\bibliographystyle{splncsnat}
\bibliography{ref}

\end{document}